\documentclass[preprint,authoryear,12pt]{elsarticle}

\usepackage{amssymb}

\journal{arXiv}

\begin{document}

\begin{frontmatter}


\title{Prediction of flu epidemic activity with dynamical model based on weather forecast}
\author[EBP]{Eugene B. Postnikov\corref{cor1}}
\ead[EBP]{postnicov@gmail.com}
\author{Dmitry V. Tatarenkov}
\address{Department of Theoretical Physics, Kursk State University, Radishcheva, 33, Kursk, 305000, Russia; Tel.:+7-4712-51-66-53}
\cortext[cor1]{Corresponding author}


\begin{abstract}
The seasonality of respiratory diseases (common cold,
influenza, etc.) is a   well-known
phenomenon studied from ancient times. The
development of predictive models is still not only an actual unsolved
problem of mathematical epidemiology but also is very
important for еру safety of public health. Here we
show that SIRS (Susceptible-Infected-Recovered-Susceptible) model
accurately enough reproduces real curves of flu activity. It
contains variable reaction rate, which is a function of mean daily
temperature.

The proposed alternation of variables represents SIRS equations as
the second-order ODE with an outer excitation. It reveals an origin
of such predictive efficiency and explains analytically the $1:1$ dynamical
resonance, which is known as a crucial property of epidemic
behavior. Our work opens the perspectives for the development of
instant short-time prediction of a normal level of flu activity
based on the weather forecast, and allow to estimate a current
epidemic level more precisely. The latter fact is based on the explicit
difference between the expected weather-based activity and
instant anomalies.
\end{abstract}

\begin{keyword}

Mathematical epidemiology, kinetics, SIRS

\MSC 92D30 \sep 34L30 \sep 62M20

\end{keyword}

\end{frontmatter}


\section{Introduction}

Seasonable variability of respiratory deceases such
as flu or common cold is widespread and generally known
phenomenon. It has been studied extensively in the last decade both
theoretically and experimentally since complications of such ``easy''
deceases result in the pneumonia, pleurisy or death. There is a big
trend in the mathematical epidemiology to develop models which
allow to reproduce in details observed phenomenon 
(\cite{Altizer2006,Fisman2007,Lofgren2007,Tamerius2011}) but it is
not enough because the prediction of outbreaks  is
more important in this case.

The origin of seasonal epidemic variability has several reasons:
social  (seasonal variations in a contact rate, e.g. schooltime
year schedule, which is important for a study of epidemics among
children) (\cite{Altizer2006,Tamerius2011}) as well as meteorological
(air temperature, humidity, illumination) conditions.

The comparative study of three factors such as
minimal day temperature, relative  humidity and cloud cover
influencing respiratory diseases reflected in the Netherlands data (\cite{Meerhof2009}) shows
prevalent correlation with the minimal temperature. The presence
of the correlation between this temperature and the relative
humidity time series without a significant lag, the fact, which is
natural for middle latitudes of Northern hemisphere, has been also
demonstrated.

The statistical analysis evaluated recently with the epidemic data
from Japan (\cite{Shojl2011}), Belgium, Portugal and the
Netherlands (\cite{Noort2012})  also does not show statistical
dominance of absolute humidity over temperature as a controlling
factor. Thus, one can choose only one of these two outer
independent variables as a free parameter. Further,
a temperature will be considered as this variable.

From the  point of view of mathematical modeling of such systems, the
mentioned seasonal  effects are simulated by introduction of
periodic coefficients into compartmental epidemiological models.
However, the corresponding studies were devoted only to the
stability analysis of mean basic reproduction number and abstract
mathematical consideration of fixed points stability 
(\cite{Williams1997,Rebelo2012}).

Some realistic models have been developed only within the last years.
It can be noted minimal  SIR (Susceptible-Infected-Recovered)
model (\cite{Noort2012}) with seasonal renewal of susceptible hosts,
which is verified by the comparison with real epidemiological
data. This kind of seasonality (resetting of variables) was
interpreted  there as a result of relatively long-term factors, e.g. effect of
antigenic drift and shift in viruses as well as vital dynamics in a population.

More realistic in the sense of continuity of the process is the
SIRS  (Susceptible-Infected-Recovered-Susceptible) model, which
connects three kinds of individuals, susceptible ($S$), infected
($I$), and recovered (and temporally immune) ($R$) within the
kinetic scheme
\begin{eqnarray}
S+I&\stackrel{k(t)}{\rightarrow}&2I\label{S}\\
I&\stackrel{\tau^{-1}}{\rightarrow}&R\label{I}\\
R&\stackrel{\theta^{-1}}{\rightarrow}&S\label{R},
\end{eqnarray}
where $k(t)$ is a reaction rate, which could be time-dependent due
to  temperature variations; and there are characteristic times
$\tau$ and $\theta$ describing the duration of illness and the
period of temporal immunity.

 The majority of works exploring epidermic oscillations within SIRS scheme deal with a probabilistic approach.
The important result of stochastic (Markov chains) modeling of a
temporal  evolution (\ref{I})--(\ref{R}) is the finding that large oscillations in the system are generated by
the $1:1$ dynamical resonance of intrinsic oscillations and
sinusoidal oscillations of the parameter $k(t)$ (\cite{Dushoff2004}).
Further, this fact was widely discussed in the context of a
possible background for annual resonant forcing of epidemic
oscillations (\cite{Fisman2007,Lofgren2007}).

Particularly, the semiannual variation of $k(t)$ (winter/summer
mean temperature), which replaces harmonic function
used in (\cite{Dushoff2004}), satisfactory reproduces characteristic
seasonal dynamics of influenza-like illness (ILI). It has been
demonstrated in the recent study (\cite{Hooten2011}) based on
the stochastic agent-based SIRS simulation and Bayesian framework for
generating of  resulting distributions, which were compared with
USA data extracted from \verb"Google" \verb"Flu" \verb"trends".

However, the cited results of stochastic simulations remain, in
principle,  phenomenological observations but do not explain the
origin of such resonant behaviour. Besides, in these papers,
artificial or too averaged parameter variations have been
considered that does not allow to analyze detailed interplay between
microstructure of flu and temperature time series. Thus, the main
goals of the present work are: i) to study analytically the
mathematical background for the dynamical resonance basing on the
ODE system corresponding to the SIRS kinetic
scheme~(\ref{S})--(\ref{R}) and ii) to demonstrate its potential
for prediction of  detailed flu activity using actual
meteorological data.

\section{Kinetic model}

The system of ordinary differential equations, corresponding to the kinetic scheme~(\ref{S})--(\ref{R}) reads
\begin{eqnarray}
\frac{dS}{dt}&=&-k(t)IS+\theta^{-1}R,\label{eqS} \\
\frac{dI}{dt}&=&k(t)IS-\tau^{-1}I,\label{eqI}\\
\frac{dR}{dt}&=&\tau^{-1}I-\theta^{-1}R. \label{eqR}
\end{eqnarray}

Let us consider the simplest way to incorporate seasonal
temperature variations $T(t)$ into  the reaction rate
$k(t)=k_0(1+\kappa(t))$, namely, we define the variable part as
the linear function $\kappa(t)=\kappa_0T(t)/|T|_{max}$, where the
positive constant $\kappa_0<1$.

To explore an ability of this approach to understand and predict
real seasonal variations in the  disease activity of ILI, we use
data from \verb"Google Flu Trends" via
http://www.google.org/flutrends/, which are argued as a valid
source for estimating flu activity (\cite{Ginsberg2009}). Note that
it was also confirmed by the study of individual outbreaks by
stochastic SIRS model (\cite{Hooten2011}). To avoid averaging over
provinces, which are presented in this database, we consider data
on flu activity in Berlin and Vienna. 
Thus, we consider data related to the averaging over the large cities, i.e. include the mixing of large well-localized population.

Both cities belong to European regions with the continental climate. The most typical example of influenza-like diseases is a common cold, which is a sufficiently continent-wide endemic seasonal illness. Therefore, a possible input into the experimental data of atypical flu-like diseases (such as new flu strains) could be negligible and ways of their geographical large-scale spread are out goals of this work. Thus, it is more preferable to neglect by the space variable and only to consider ODE equations for the modeling. 

\begin{figure}
\includegraphics[width=\columnwidth]{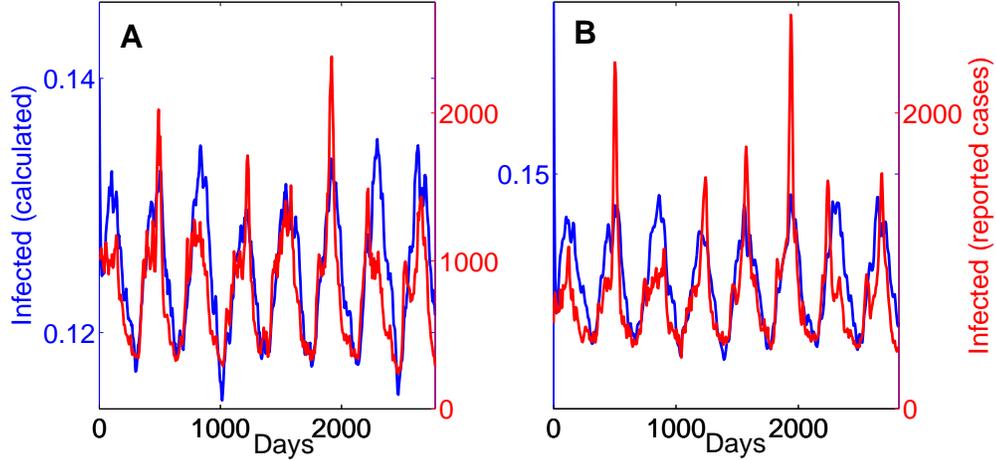}
\caption{(Color online) Comparison of time series for calculated
(dark-gray (blue online) lines)  infected part of full population
normed by unity and corresponding number of reported cases
(light-gray (red online) lines) for Berlin (A) and Vienna (B). For
both cities parameters of the model (\ref{eqS})--(\ref{eqR}) are
equal: $k_0=0.21$, $\kappa_0=-0.07$, $\tau=7$, $\theta=10$; time
intervals: October, 19, 2003 till May 22, 2011.}
\label{BerlinWien}
\end{figure}

The  daily mean temperature
data series are taken from \verb"European Climate"
\verb"Assessment & Dataset" (\cite{KleinTank2002})
(http://eca.knmi.nl/) for stations Berlin-Tegel (ECA station code:
4005) and Wien (Vienna, ECA station code: 16), daily sampling.
Results of numerical solutions of the system
(\ref{eqS})--(\ref{eqR}) with the substituted
temperature-dependent reaction rate mentioned above are presented
in the Fig.~\ref{BerlinWien} in comparison with data series from
\verb"Google Flu Trends".

Solution of the system (\ref{eqS})--(\ref{eqR}) was evaluated by
MATLAB R2006a  routine \verb"ode45" realizing the Runge-Kutta 4-5
method with the relative tolerance $1e-7$, temperature values between
sample's nodes were linearly interpolated. Calculations for
linearized model (\ref{filtering}) were evaluated using MATLAB
routines for the Fast Fourier Transform.

One can see that  the numerical
simulation of the model  sufficiently good reproduces actual flu
activity for both considered samples. First of all, the simulated oscillations are undamped
and one can note the high accuracy of coincidence of periods for
measured and simulated curves. This clearly indicates that the
used temperature variations enforce the desired
oscillations.

But even more important is the coincidence of not only periods but
also shapes of curves,  especially in the minima of each period's
magnitude. One can clearly see that even small saw-like details
overlap for many lower parts of both real and simulated curves in
the Fig.~\ref{BerlinWien}. 

Apart of visual comparison, let us provide some quantitative statistical results on flu data and theirs temperature-based simulation shown in the Fig.~\ref{BerlinWien}. Table~\ref{tab} presents total correlation coefficient for both curves over all period of simulations as well as correlation coefficients for each period of oscillatory flu outbreaks. To determine these short-time correlation coefficients, full time interval is subdivided into parts corresponding to unique individual periods of oscillations. Their boundary points are determined as time moment of cross-section of the curve presenting \verb"Google Flu" data and the median line for the data. The starting point for each subinterval is chosen in top-down cross-section point. As a result, the correlation coefficients are determined for experimental and calculated time series located between closest subdivision points.

\begin{table}
\label{tab}
\caption{Correlation coefficients for the pairwise real and simulated oscillating time series presented in the Fig.~\ref{BerlinWien}}
\begin{tabular}{|p{0.09\textwidth}|p{0.08\textwidth}|p{0.08\textwidth}|p{0.08\textwidth}|p{0.08\textwidth}|p{0.08\textwidth}|p{0.08\textwidth}|p{0.08\textwidth}|p{0.07\textwidth}|}
\hline
City& \multicolumn{7}{|c|}{Correlation coefficients for sequential individual periods}&	for total time \\
\hline
Berlin &0.89	&0.84	&0.82	&0.89	&0.85	&0.70	&0.77&	0.78\\
\hline
Vienna	&0.69	&0.83	&0.77	&0.83	&0.82	&0.53	&0.75	&0.70\\
\hline
\end{tabular}
\end{table}

One can see that correlation coefficient is larger within majority of these subintervals than for the distributions taken for the whole interval. This fact principally originates from the epidemic outbreak within the 6th full period of oscillations for both cities. The values of most of the other coefficients argue that not only period of oscillations are the same but and their shape is sufficiently correlated even for such simple model. This also answers the question about other seasonal factors (like school vacation): the correlation coefficient of the model temperature-dependent solution and the real data are even higher within most of individual seasons then averaged over many seasons periodicity. Correspondingly, the temperature dependence could be considered as a leading factor.

This  confirms the point of view that
there exists a natural level of seasonal flu activity, which is
principally determined by the variation of a daily mean air
temperature.  At the intermediate level of activity, this direct
influence can also be traced: there are the same shape details,
however some time they shifted relative to each other. These
displacements can be explained by the presence of irregular
temporary localized outbreaks, which depend on a variety of factors
(social, virological, etc).  These factors  play a major role at
highest levels of flu activity. That is why, the several seasonal
maxima  differ from weather-based prediction  being connected with
an actual epidemiological situation.

To reveal the origin of a resonant behaviour, let us apply the
co-ordinate transformation  providing more explicit representation
of a forcing. First of all, the system (\ref{eqS})--(\ref{eqR}) is
actually two-dimensional due to the conservation law $S+I+R=1$. It
is more convenient to introduce new variable
$N=\tau^{-1}I-\theta^{-1}R$ that extremely reduces the
Eq.~(\ref{eqR}), denoting $N$ simply as a time derivative of $R$.

For the constant reaction rate $k=k_0=\mathrm{const}$ the standard
procedure shows  that the ODE system with respect to variables
$(N,R)$ has a non-trivial stationary point
\begin{equation}
N_s=0,\quad R_s=\frac{1-\tau^{-1}k^{-1}_0}{1+\tau\theta^{-1}}.
\label{Rs}
\end{equation}
Linear analysis with respect to $N$, $r=R-R_s$ declares that
$(0,\,R_s)$ is a stable  focus. In other words, any small
deviation from $(0,\,R_s)$ decays oscillatory with the intrinsic
frequency $\omega_i=[\omega_0^2-(\lambda/4)^2]^{1/2}$, where
$\lambda=\tau^{-1}+\theta^{-1}+R_sk_0\left[1+2\tau\theta^{-1}\right]-k_0$
and $\omega_0=\theta^{-1/2}\left(k_0-\tau^{-1}\right)^{1/2}$.

Thus, considering a general time-dependent reaction rate
$k(t)=k_0[1+\kappa(t)]$, $|\kappa|<1$, it is possible to
represent the Eqs.~(\ref{eqR})--(\ref{eqS}) in the form

\begin{eqnarray}
\frac{dr}{dt}&=&N, \label{eqrt}\\
\frac{dN}{dt}&=&
R_s\theta^{-1}\tau^{-1}\kappa(t)-\nonumber\\
&&\left(\tau^{-1}+\theta^{-1}+R_sk(t)\left[1+2\tau\theta^{-1}\right]-k(t)\right)N-
\theta^{-1}\left(k(t)-\tau^{-1}\right)r- \nonumber\\
&&k(t)\tau N^2-k(t)(1+\tau\theta^{-1})Nr-k(t)\theta^{-1}\left[1+\tau\theta^{-1}\right]r^2. \label{eqnt}
\end{eqnarray}

This form allows to discuss influence of a reaction rate
variability in  a most explicit way. First of all, the first term
in the left-hand side of the Eq.~(\ref{eqnt}) demonstrates that the
variability of the reaction  rate corresponds actually to an
external time-dependent excitation. This means that resonant
properties of the solution for SIRS equations  have
primarily outer-, not parametric resonance origin in this case. And this
result accomplishes for perturbations arbitrary strength and
functional character (the case of small harmonic $\kappa(t)$ will
be considered and discussed below).

The linear (second line in the Eq.~(\ref{eqnt})) and non-linear
(third line in the Eq.~(\ref{eqnt})) terms include  time-varying
$k(t)$ as well. Therefore, in the case of sufficiently large
$\kappa(t)$, some parametric excitation could be detected too.
However, due to positivity of fixed parameters and their range
resulting in both positivity of $R_s$ and strong decay of free
outbreaks, these inner effects are as a rule smaller then outer
ones.

Finally, the realistic conditions $N<<1$, $r<<1$ (number of
individuals on various stages of illness is sufficiently  smaller
than a full population) provides an opportunity to neglect by
non-linear terms in the Eq.~(\ref{eqnt}) for the case of steady
oscillatory regime and for $|\kappa|<<1$ to neglect its variation
in the second line of the Eq.~(\ref{eqnt}). Under these
assumptions, the system (\ref{eqrt})--(\ref{eqnt}) reduces to the
simple standard linear second order ordinary differential
equation. It should be noted that the mathematically similar
system has been obtained by K.~Dietz (\cite{Dietz1976}) in the
problem of another epidemiological origin: endogenous diseases
with vital dynamics and restriction to the simple harmonic
variation of reaction rate.

For this reason, we neglect by the non-linear terms (the third
line in the Eq.~(\ref{eqnt})) and variable  parts of $k(t)$ in the
linear ones (the second line in the Eq.~(\ref{eqnt})) as small
quantities. Thus, the rest expression is a simple non-homogeneous
harmonic ODE and the resulting solution for the number of infected
individuals reads as
\begin{equation}
I=\tau\theta^{-1}R_s+
\kappa_0\frac{R_s\theta^{-1}}{2\pi}\int_{-\infty}^{+\infty}\frac{\theta^{-1}+i\omega}
{\left[\omega_0^2-\omega^2\right]+i\omega\lambda}\hat{\kappa}(\omega)d\omega,
\label{filtering}
\end{equation}
where $\hat{\kappa}(\omega)$ is a Fourier transform of time series for a normed mean temperature.

\begin{figure}
\includegraphics[width=\columnwidth]{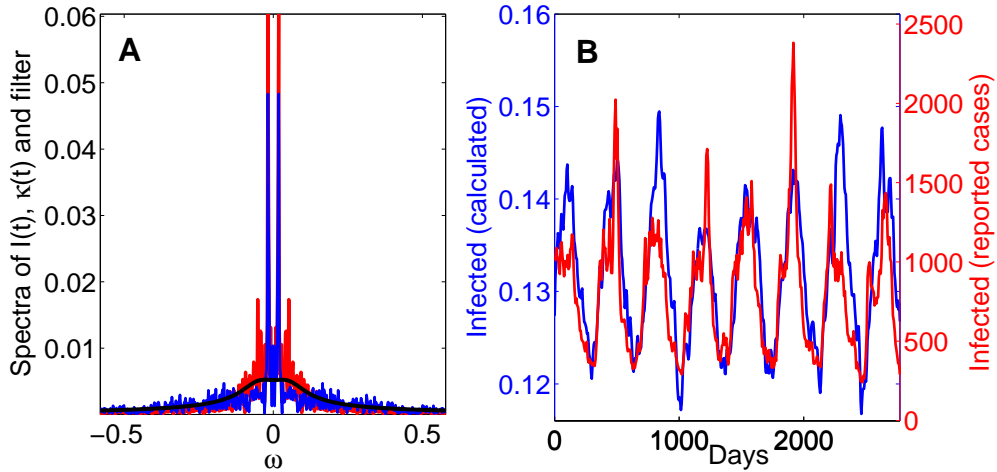}
\caption{(Color online) A) Normalized to unit area under curves,
absolute values for spectra of  daily mean temperature (dark-gray
(blue online) curve), reported flu activity (light-gray (red
online) curve) and resonant filter (black curve). B) Comparison of
calculated via the expression~(\ref{filtering}) (dark-gray (blue
online) curve) and actually reported (light-gray (red online)
curve) common flu activity in Berlin. The parameters of the model
are the same as in the Fig.~\ref{BerlinWien}.} \label{lin}
\end{figure}

The formula (\ref{filtering}) explains the phenomena of $1:1$
dynamical resonant  excitation for sinusiodal (\cite{Dushoff2004})
and stochastic ({\cite{Black2010}}) perturbations of the reaction rate
since the integral kernel has a form of resonant filter, which
amplify the spectral component coinciding with the intrinsic
frequency of oscillations. In more general case, this kernel
reshapes time series of temperature oscillations into time series
of flu activity. The illustration of this procedure is presented
in the Fig.~\ref{lin}.

Considering Fig.~\ref{lin}A, one can see that the spectrum of
reported flu activity has  two prevailing peaks corresponding to
the first and second harmonics of time series and relatively fast
decay elsewhere. The temperature spectrum has main
harmonics located in the same place (however with smaller amplitude) and a thicker tail,
which take over a significant part of spectral density. Thus, the
harmonic filter in the Eq.~(\ref{filtering}), which is drawn as
the black line in the Fig.~\ref{lin}, boosts the central part of
temperature spectrum providing its better coincidence with the flu
activity spectrum. This results in a rather good reproduction of flu
activity curve, see Fig.~\ref{lin}B even in this simplified
linearized case. Moreover, comparison of
Figs~\ref{lin}B~and~\ref{BerlinWien}A shows their high similarity
in a shape. Thus, this indicates that harmonic filtering
(\ref{filtering}) is the actual governing factor, which determines
seasonal flu activity through temperature seasonality.

\section{Summary and outlooks}

In this work, we show that seasonal mean temperature variation
determines principal activity of influenza-like diseases and the
latter can be calculated basing on the classical SIRS model  with
the variable temperature-dependent reaction rate.

The proposed method can be used for an instant short-time
prediction of local level influenza-like illnesses basing on current
epidemiological level and weather forecast. Current activity data
 (instant short-time sample) allow
to adjust parameters of SIRS ODE system and initial conditions.
This equation should be solved numerically with the variable
reaction rate, which incorporates temperature time series obtained from a
weather forecast. The result provides prediction of a ``normal''
(weather-based) flu activity for nearest days (up to a decade). It
should be pointed out that a definition of this ``normal level of
epidemiological activity'' is the question, which is discussed
from times of birth modern mathematical epidemiology, see
e.g. (\cite{Hedrich1927}) up to now, see e.g. the
review (\cite{Stephenson2002}).

Our approach allows to refine this definition via a replacement of pure averaging over a long time interval with
the instant value. This will  allow avoid misinterpretation of a
danger of new respiratory virus strain since the instant temperature anomalies can
result in lower or higher levels of outbreak.
Anomalies calculated as a difference between observed and expected
temperature-based flu activity will give more reasoned alarms.

Thus, the proposed model has a predictive power and opens perspectives for future detailed research for a variety of world's regions. Further, it allows developing technological (say, web-based) forecast applications. 

Finally, it should be noted that  the studied system belongs to the wide class of ODEs applicable to various problems of physical chemistry and biophysics. Thus, the obtained results could be used  for the search of new approaches to a parametric control in autocatalytic systems that is a permanent interest of non-linear dynamics.


\end{document}